\documentstyle[12pt]{article}
\input epsf

\begin{document}

\begin{titlepage}

\begin{flushright}
UCLA-00-TEP-7\\
UM-TH-00-03\\
February 2000
\end{flushright}
\vspace{2.0cm}

\begin{center}
\large\bf
{\LARGE\bf Exact ${\cal O}(g^2 \alpha_s)$ top decay width
                      from general massive two-loop integrals}\\[2cm]
\rm
{Adrian Ghinculov$^a$ and York-Peng Yao$^b$}\\[.5cm]

{\em $^a$Department of Physics and Astronomy, UCLA,}\\
      {\em Los Angeles, California 90095-1547, USA}\\[.1cm]
{\em $^b$Randall Laboratory of Physics, University of Michigan,}\\
      {\em Ann Arbor, Michigan 48109-1120, USA}\\[3.0cm]
      
\end{center}
\normalsize

\begin{abstract}
We calculate  the
$b$-dependent self-energy of the top quark at ${\cal O}(g^2 \alpha_s)$
by using a general massive two-loop algorithm proposed in a previous article. 
From this we derive by unitarity the ${\cal O}(\alpha_s)$ radiative 
corrections to the decay width of the top quark, where all effects
associated with the $b$ quark mass are included without resorting to
a mass expansion. Our results
agree with the analytical results available for the 
${\cal O}(\alpha_s)$ correction to the top quark width.
\end{abstract}

\vspace{3cm}

\end{titlepage}


\title{Exact ${\cal O}(g^2 \alpha_s)$ top decay width
                      from general massive two-loop integrals}

\author{Adrian Ghinculov$^a$ and York-Peng Yao$^b$}

\date{{\em $^a$Department of Physics and Astronomy, UCLA,}\\
      {\em Los Angeles, California 90095-1547, USA}\\
      {\em $^b$Randall Laboratory of Physics, University of Michigan,}\\
      {\em Ann Arbor, Michigan 48109-1120, USA}}

\maketitle

\begin{abstract}
We calculate  the
$b$-dependent self-energy of the top quark at ${\cal O}(g^2 \alpha_s)$
by using a general massive two-loop algorithm proposed in a previous article.
From this we derive by unitarity the ${\cal O}(\alpha_s)$ radiative 
corrections to the decay width of the top quark, where all effects
associated with the $b$ quark mass are included without resorting 
to a mass expansion. Our results
agree with the analytical results available for the 
${\cal O}(\alpha_s)$ correction to the top quark width.
\end{abstract}


Precision measurements of the electroweak parameters are a powerful tool
for testing the validity of the standard model and searching for new physics.
The impressive accuracy attained experimentally, which is also expected to
improve in the future, makes a complete two-loop analysis of the data 
necessary. 

However, on the theoretical side, such a complete two-loop
electroweak analysis is far from being a simple task. Apart from  
the proliferation of diagrams, a special kind of technical problem is 
encountered in electroweak two-loop calculations. These processes 
involve particles with different masses, and in general need to be 
evaluated at finite external momenta. It has been known already for
some time that general massive two-loop Feynman diagrams,
when evaluated at nonvanishing external momenta, lead to complicated
and often unknown special functions. See, for instance, ref. \cite{lauricella}
which relates the sunset two-point topology to the Lauricella function.

Recent progress in two-loop electroweak analyses was mainly attained by
using mass or momentum expansion methods. 
We note that in some situations mass expansions
turn out to converge well, such as the top mass expansion of the $\alpha_s$
corrections to the $Z \rightarrow b \bar{b}$ process \cite{topexpansion}, 
while in other cases subleading terms are known to be 
substantial \cite{degrassi}. 
In certain situations it was possible to recover the exact function starting 
from an expansion \cite{ritbergen}. On the other hand, when the process under 
consideration involves more than one ratio of masses, recovering the 
exact result from an expansion would be difficult, the functions involved
being complicated.

In order to circumvent the use of a mass or momentum expansion, we
proposed in ref. \cite{2loopgeneral} a general framework for treating two-loop
Feynman graphs by a combination of analytical and numerical methods.
Given a Feynman diagram, its tensorial structure is first reduced
analytically into a combination of ten special functions $h_i$, which are
defined by fairly simple integral representations. The result of this
analytical reduction is further integrated numerically. 

A subset of the ten special functions $h_i$ \cite{2loopgeneral},
namely $h_1$ and $h_2$, are sufficient for treating two-loop corrections
in a theory involving only scalars and no derivative couplings. This is
the case of radiative corrections of enhanced electroweak strength in
the standard model, and in ref. \cite{2loop} this method was used for deriving
corrections of leading power in the Higgs mass. The resulting radiative 
corrections agree with independent calculations which use different 
techniques \cite{kniehl}.

Going from a scalar theory to a full renormalizable theory with 
fermions and derivative couplings, such as the standard model,
increases a lot the complexity of the problem. The analytical 
reduction of the tensor structure of a graph into $h_i$
functions along the lines of ref. \cite{2loopgeneral}
needs to be handled by a symbolic manipulation program 
such as FORM or Schoonschip, and results in rather lengthy
expressions. Once the reduction is performed, the numerical
integration methods are the same as those used in ref. \cite{2loop},
with the only difference that more computing power is needed to
handle the number of $h_i$ functions which need to be integrated.

In ref. \cite{expansion} we have shown how these methods 
can be used for calculating
momentum derivatives of two-loop two-point functions around a finite, 
non-zero value of the external momentum. Such momentum 
derivatives are encountered for instance when evaluating 
wave function renormalization constants on-shell. 

In this letter we show how this method can be used for calculating
two-point functions of physical interest. We note that, due to the
general nature of the tensor reduction algorithm given in 
ref. \cite{2loopgeneral},
the inclusion of more than two external legs proceeds in the same way.
The difference is that the resulting expressions are more complicated
than those stemming from two-point functions, and more computing power
is needed for performing higher-dimensional numerical integration.

\begin{figure}[t]
\hspace{1.cm}
    \epsfxsize = 14.5cm
    \epsffile{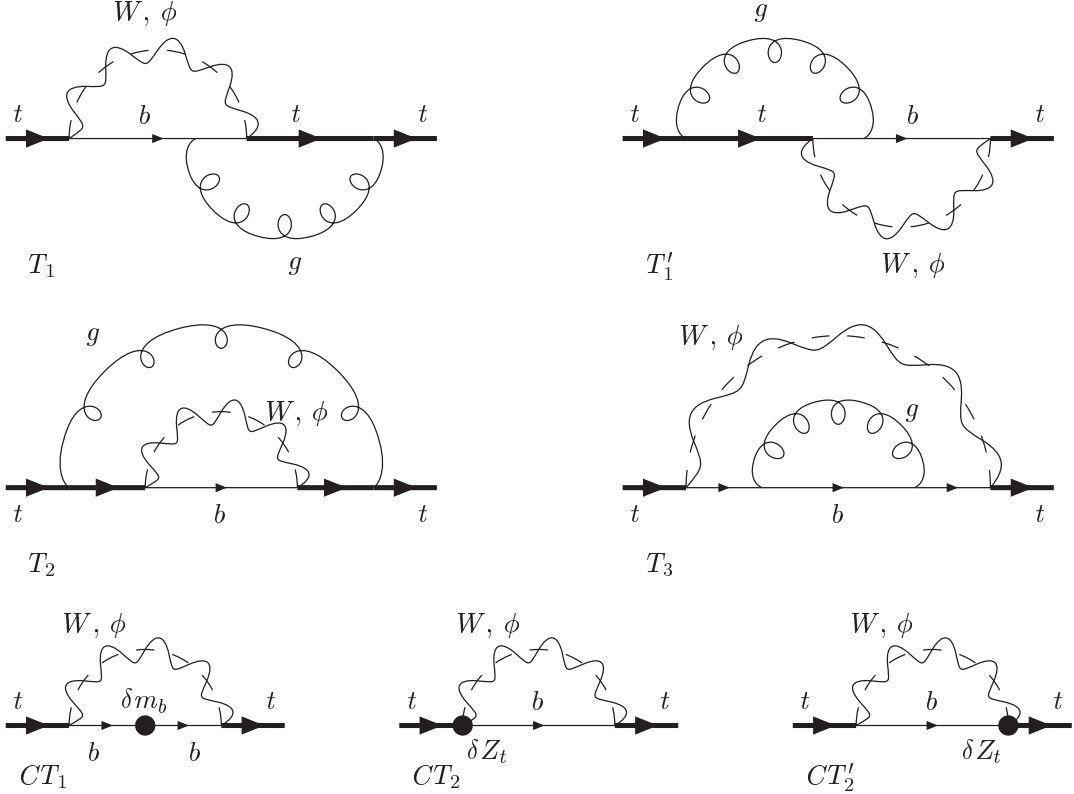}
\caption{{\em The two-loop Feynman graphs which
              contribute to the $b$-mass dependent correction 
              of ${\cal O}(\alpha_s g^2)$ to the top self-energy.
              Only the counterterm diagrams are shown which are needed
              for subtracting the infinities of the imaginary part of the
              self-energy, which gives the ${\cal O}(\alpha_s)$ correction
              to the $t \rightarrow W+b$ decay.}}
\end{figure}

In figure 1 we show the two-loop Feynman graphs relevant for the
$b$-dependent self-energy of the top quark at order $g^2 \alpha_s$.

As for the counterterm structure, we only show in figure 1 the 
counterterm diagrams which are necessary for renormalizing the imaginary
part of the self-energy. These are the on-shell one-loop QCD counterterms of
the $b$-quark mass $\delta m_b$ and of the top wave function renormalization
constant $\delta Z_t$:

\begin{eqnarray}
  \delta m_b & = &  2 \alpha_s N_c \pi^2 m_b
     \left\{ \frac{3}{\epsilon} 
           + \frac{3}{2} \left[ \gamma + \log{\pi} 
           + \log{\left(\frac{m_b^2}{\mu^2}\right)} \right] - 2 \right\}
 \\ \nonumber
  \delta Z_t & = & \alpha_s N_c \pi^2  
     \left[ \frac{2}{\epsilon} + \gamma + \log{\pi} 
                               + \log{\left(\frac{m_t^2}{\mu^2}\right)} 
                               + 2 \log{\left(\frac{m_t^2}{m_g^2}\right)} 
                               - 4 \right]
\end{eqnarray}
Here, $\mu$ is the 't Hooft mass scale, and $m_g$ is the gluon mass 
infrared regulator. $N_c=4/3$ is the color factor.

The imaginary part of the two-loop self-energy on-shell 
is physical. Writing the top self-energy as 
$\Sigma(p \cdot \gamma)=\Sigma_1(p \cdot \gamma) 
+ \gamma_5 \cdot \Sigma_{\gamma_5}(p \cdot \gamma)$,
the top decay width is given by $\Gamma_t=2 \cdot Im 
\Sigma_1(p \cdot \gamma = m_t)$.
This can be compared to known analytical and numerical results for the 
${\cal O}(\alpha_s)$ QCD corrections to the top decay 
width \cite{topdec1loop}, 
and thus provides a nontrivial test of the two-loop algorithm.

\begin{figure}
\hspace{1.cm}
    \epsfxsize = 14.5cm
    \epsffile{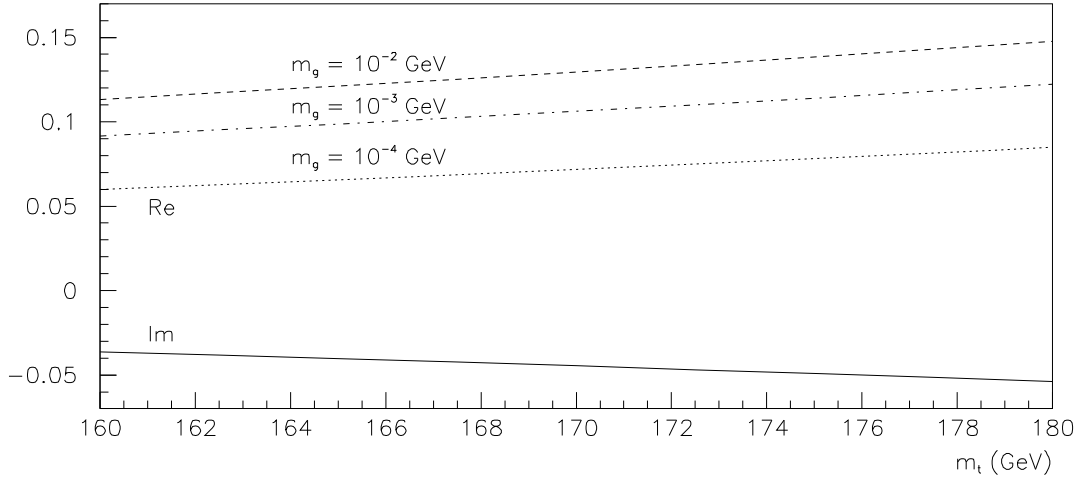}
\caption{{\em The real and the imaginary parts of 
              $\Sigma_1(p \cdot \gamma = m_t)$ at two-loop
              (see text), given by the graphs of figure 1, 
              as a function of the top mass and of the gluon infrared
              mass regulator. An overall factor 
              $\alpha_s  N_c  G_F  M_W^2  m_t /\sqrt{2} /(2 \pi)^8$
              is understood.}}
\end{figure}

The algebraic reduction of the tensorial structure of the graphs 
was done by implementing the reduction formulae of 
ref. \cite{2loopgeneral} into
a symbolic manipulation program. The resulting intermediary decomposition into
$h_i$ functions is too lengthy to reproduce here. Instead,
we give the final results obtained after numerical integration over the
analytical decomposition.

Here we would like to discuss briefly the treatment of infrared
divergence associated with the massless gluon.
To regularize the infrared singularities,
we introduce a mass regulator for the gluon.
We note that in higher-order QCD calculations introducing a 
gluon mass would affect the Slavnov-Taylor identities.
In such a case, a different approach is needed to treat the
infrared singular diagrams. One possibility is to try to extract
analytically the IR singular part of the diagram and to treat
this by dimensional regularization, while the remaining IR finite
part of the diagram can be calculated by numerical integration. 
Whether or not this can always be done in a simple way is beyond
the scope of this article. We also note that our approach is mainly designed
to treat massive graphs, while other approaches are available for 
massless QCD calculations.  However, this being an 
${\cal O}(g^2 \alpha_s)$ correction, the use of a gluon mass 
regulator is legitimate in this case.
The physical quantity $Im \Sigma_1(p \cdot \gamma = m_t)$,
which is associated with the top quark width, is free of infrared 
singularities, and we checked numerically that indeed the 
result becomes independent of the gluon regulator when the
mass regulator is much smaller than the $b$ mass.

In figure 2 we give numerical results for the ultraviolet finite 
part of the quantity $\Sigma_1(p \cdot \gamma = m_t)$ derived from 
figure 1; the inclusion of the 
counterterm contributions shown in figure 1 makes only the 
imaginary part (which is physical) of the self-energy finite, both UV and IR. 
We plot the results for a range
of the top mass, and assume $G_F=1.16637 \cdot 10^{-5}$ GeV$^{-2}$, 
$m_W=80.41$ GeV, $m_b=4.7$ GeV, and $\alpha_s(m_t)=.108$.

\begin{table}
\begin{tabular}{||c||c|c|c|c|c||}                                         \hline\hline
 $m_t$ [GeV]                                      &  160  &  165  &  170  & 175   & 180    \\ \hline\hline
 $\Gamma_t^{tree}$ [GeV]           & 1.127 & 1.260 & 1.402 & 1.553 & 1.712  \\ \hline
 $\delta \Gamma_t^{1-loop}$ [GeV]  & -.092 & -.104 & -.117 & -.132 & -.149  \\ \hline\hline
\end{tabular}
\caption{The ${\ O}(\alpha_s)$ correction to the top decay  
$t \rightarrow W+b$ as obtained from the imaginary part of
the two-loop top self-energy of figure 1, integrated numerically. 
We took $G_F=1.16637 \cdot 10^{-5}$ GeV$^{-2}$, 
$m_W=80.41$ GeV, $m_b=4.7$ GeV, and $\alpha_s(m_t)=.108$.}
\end{table}

In table 1 we give numerical values for the ${\cal O}(\alpha_s)$ correction
to the top decay rate $t \rightarrow W+b$ obtained from the imaginary
part of the self-energy. Therefore, the ${\cal O}(\alpha_s)$ QCD 
correction is obtained as an inclusive quantity, 
integrated over the gluon spectrum. To check the size of the finite mass
of the $b$ quark, we ran our programs in the vanishing $b$ mass limit.
At tree level the finite $b$ mass amounts to a 3-4 MeV correction to the width,
and in the ${\cal O}(\alpha_s)$ correction the effect is negligible.
These results agree with existing calculations of 
${\cal O}(\alpha_s)$ corrections to $t \rightarrow W+b$ \cite{topdec1loop}, 
which provides a good test for the two-loop tensor decomposition and  
numerical integration algorithm.

To conclude, we have shown that the general methods proposed in 
ref. \cite{2loopgeneral} 
can be used to calculate physical radiative corrections. We treated 
at two-loop order the $b$-mass dependent self-energy of the top quark 
at ${\cal O}(g^2 \alpha_s)$. From the imaginary part of the self-energy we
extract the ${\cal O}(\alpha_s)$ corrections to the top decay process
$t \rightarrow b + W$, and find agreement with existing results for this 
process. The calculation is performed while respecting the actual mass and
momentum kinematics of the process, and without resorting to a mass or 
momentum expansion of the diagrams. 

\vspace{1cm}

{\bf Aknowledgements}

The work of A.G. was supported by the US Department of Energy.
The work of Y.-P. Y. was supported partly by the US Department of Energy.



\end{document}